\documentstyle[twoside,fleqn,espcrc2]{article}
\input{epsf}

\newcommand{\EQ}{\begin{equation}}
\newcommand{\EN}{\end{equation}}
\newcommand{\AmS}{{\protect\the\textfont2
  A\kern-.1667em\lower.5ex\hbox{M}\kern-.125emS}}

\title{Glueball wavefunctions}

\author{\underbar{Philippe de Forcrand}\address{IPS, ETH Zentrum,
        CH-8092 Z\"urich, Switzerland}
        and
        Keh-Fei Liu\address{Dept. of Physics and Astronomy, Univ. of Kentucky,
        Lexington, KY 40506}
}

\begin{document}

\begin{abstract}
We measure Coulomb-gauge wavefunctions of the scalar and tensor glueballs
in SU(2) Yang-Mills.  The problem of contamination by flux states is discussed,
and a new analysis method described.  The large size
of the tensor glueball is confirmed.  Preliminary results in the deconfined
phase show no significant changes.
\end{abstract}

\maketitle


Little is known about pure Yang-Mills glueballs besides their masses.  Precise
lattice
estimates of their size, form factor, matrix elements would help considerably
their experimental search.  Meanwhile the study of lattice observables in a
fixed gauge (gluon, quark propagator, baryon and meson wavefunctions) is
receiving a lot of attention.  If only for comparison, it is
desirable to measure glueball wavefunctions in Coulomb gauge.  In addition,
gauge-fixing allows the construction of operators with a reduced degree of
ultraviolet divergence, which compare well with gauge-invariant operators for
measuring physical quantities like glueball masses.

In previous work \cite{PRL}, we used a static source method to create
ground-state glueballs,
and extracted from lattice simulations a form factor for the tensor.
Here we measure directly correlations of Coulomb-gauge glueball operators, thus
obtaining information on excited states as well.  We present a new method of
analysis to remove flux state (also called torelon) contributions.  And we
report preliminary results at finite temperature.

\section{Lattice observable}

We work with SU(2) and with the Wilson action.  At regular intervals (50
sweeps)
in the gauge-invariant Monte Carlo process, we fix the gauge to Coulomb
by maximizing
\EQ
\sum_{\vec{x}} \sum_{\mu = 1, 3} Re Tr U_{\mu}(\vec{x},t)
\EN
for every time-slice $t$, using over-relaxation.  The issue of Gribov copies is
not addressed here.  We think that errors due to this neglect are small
compared
to our other errors.
Our lattice observable is then constructed in two stages.

Decomposing an SU(2) link matrix
\EQ
U_{\mu}(\vec{x}) = c_0(\vec{x},\mu) + i \sum_{i=1,3} c_i(\vec{x},\mu) \sigma_i
\EN
we construct the $\vec{0}$-momentum expression
\EQ
\tilde{C}_{\mu \nu}(\vec{r},t) \equiv \sum_{\vec{x}} \sum_{i=1,3}
c_i(\vec{x},\mu)
c_i(\vec{x}+\vec{r},\nu)
\EN
which reduces to
$a^2 \sum_{\vec{x}} A_{\mu}(\vec{x}) A_{\nu}(\vec{x} + \vec{r}) + {\cal O}
(a^4)$
\linebreak
as $a \rightarrow 0$. $\tilde{C}_{\mu \nu}$ measures the spatial connected
correlation of the gluon field.

Given a pair of spin- and orbit- quantum numbers $S$ and $L$ for the desired
glueball, $\tilde{C}_{\mu \nu}$'s can be combined via a lattice
version of
the convolution with spherical harmonics $Y_{St}(\hat{\mu})$,
$Y_{Lm}(\hat{r})$.
We form the scalar $L = S = 0$ combination
\EQ
W_{0}(|\vec{r}|,t) = \sum_{\mu} \sum_{\hat{r}} \tilde{C}_{\mu\mu}(\vec{r},t)
\EN
and the tensor $L = 0, S = 2$ combination
\EQ
W_{2}(|\vec{r}|,t) = \sum_{\hat{r}} \lbrack 2 \tilde{C}_{33}(\vec{r},t)
- \tilde{C}_{11}(\vec{r},t) - \tilde{C}_{22}(\vec{r},t) \rbrack
\EN
Note that we selected the $E$ representation of spin 2 on the lattice.
Choosing the alternative $T_2$ representation would give:
\EQ
W'_{2}(|\vec{r}|,t) = \sum_{\hat{r}} \tilde{C}_{12}(\vec{r},t)
\EN

The 3 degenerate $T_2$ states and the 2 degenerate $E$ states combine, as
$a \rightarrow 0$, to form the 5 states of a $^{5}S_2$ tensor glueball.
$^{3}P_2$ and $^{1}D_2$ states can also be
constructed with a similar prescription, for a complete spectroscopic study
of the tensor glueball.  Our first results indicate that such states are
heavier than the $L = 0$ states we study, confirming the expected energy
cost of orbital excitations.

Correlation matrices of $W$'s are then measured
\EQ
C_{ij}(t) = \newline
\langle W(r_i,0) W(r_j,t)\rangle - \langle W(r_i)\rangle  \langle W(r_j)\rangle
\EN
If $\psi^{(k)}(r)$ is the radial wavefunction of the $k$-th eigenstate of the
transfer matrix, then
\EQ
C_{ij}(t) = \sum_k \alpha_k \psi^{(k)}(r_i) \psi^{(k)}(r_j) e^{- m_k t}
\EN
The extraction of glueball masses $m_k$ and wavefunctions $\psi^{(k)}(r)$
from Monte Carlo averages of $C_{ij}(t)$ is described in the next section.

\begin{figure}[tb]
\epsfxsize=65truemm
\centerline{\epsffile[0 130 750 530]{fig1.ps}}
\caption{Correlations of $W_{0}$ (eq.4) for the three lightest states.
 The lattice is $16^4$. The dotted line shows correlations of
 fuzzy loop measured on the same sample of 180 configurations.}
\label{fig:1}
\vspace{-5mm}
\end{figure}

The overhead of measuring the $W$'s is quite small if gauge-fixing is performed
anyway to measure other gauge-fixed observables (eg. wavefunctions of ordinary
hadrons).  And, since $W$ is an ${\cal O}(a^2)$ operator only, the
signal-to-noise
ratio is quite good and compares well with fuzzing or smearing methods, while
giving additional information on excited states of the transfer matrix.
See Fig.1 for a comparison with fuzzing \cite{Teper} on the same set of
configurations.

Our decision to fix the gauge with a non-local gauge condition has a major
drawback however.  It mixes topologically different sectors.  Our link-link
observable $\tilde{C}_{\mu \nu}$ sums up a large number of loops.  Some of them
have zero winding number, as in Fig.2a, and project on glueballs.  Others
have non-zero winding number, as in Fig.2b, and project on flux states also
called torelons.  These states are not lattice artifacts, but physical states
of the finite volume spectrum.  Their mass diverges in the thermodynamic limit.
Unfortunately the torelon ground state, of mass $\sigma L a$, where $\sigma$
is the string tension and $L a$ the spatial extent of the lattice, is as
light or lighter than the ground state glueball in the accessible regime of
$\beta$ and $L$.  So the disentangling of glueball and torelon, which is
usually taken care of automatically by the choice of Wilson or Polyakov loops,
needs to be done by hand here.

\begin{figure}[tb]
\epsfxsize=65truemm
\centerline{\epsffile{fig2.ps}}
\caption{Contributions to link-link observable eq.(3) from (a) glueballs
 and (b) flux states.}
\label{fig:2}
\vspace{-5mm}
\end{figure}

\section{Dealing with flux states}

We have tried two different methods.  The results presented at LAT92 were
obtained by the first one.  The second method has been developed and tried
since the conference.

The "brute force" method (method {\bf A}) consists in diagonalizing the
correlation
matrices $C(t)$ for successive times $t$.  $C(t)$ is real symmetric so that
\EQ
C(t) = \tilde{R}(t) D(t) R(t)
\EN
where $D$ is a diagonal matrix of eigenvalues and $R$ a rotation matrix
whose columns are the eigenvectors of $C$.  We assume that we have pure
states: in the absence of statistical noise, each eigenvector of $C$
matches an eigenstate $\psi^{k}(r)$ of the complete transfer matrix (see eq.8).
This assumption looks natural and can be checked a posteriori: if several
states of distinct masses mix, this mixing will depend on Euclidean time,
and the eigenvectors of $C(t)$ will change with $t$.

Then we simply need to label each eigenvector "glueball" or "flux state".
This classification is performed by comparing masses measured at different
values of $\beta$ or different lattice sizes $L$: the physical (dimensionful)
mass $\sigma L a$ of the torelon will vary linearly with $L$ or $a$,
while that of glueballs will undergo small corrections.
We studied 3 $\beta$-values: 2.35, 2.5 and 2.7 on a $16^4$ lattice.
It turns out that the
glueball is the lightest state at $\beta = 2.35$, while the torelon is
lighter at $\beta = 2.7$ and marginally so at $\beta = 2.5$.
The situation is similar in the scalar and tensor channels.

\begin{figure}[tb]
\epsfxsize=65truemm
\centerline{\epsffile[0 130 750 530]{fig3.ps}}
\caption{Wavefunctions of tensor glueball and torelon, using method A.  The
dotted line corresponds to a time-separation $t = 0$.  Very little variation
with $t$ is apparent.}
\label{fig:3}
\vspace{-5mm}
\end{figure}

The wavefunction of the flux state turns out to be very flat, as
expected from the picture of a vibrating string.  Surprisingly however,
the ground-state glueball wavefunction possesses a
node even though the orbital angular momentum $L$ is 0.  This is shown
in Fig.3 for the tensor channel at $\beta = 2.5$.  This node persists
as the spatial size of the lattice is varied, but occurs at a radial
distance $r$ which increases with the lattice size.
In Fig.4 the scalar glueball wavefunctions at the 3 chosen $\beta$-values
are displayed together as a function of physical distance, where the
scale at each $\beta$ was set by the measured glueball mass; the
squeezing of the wavefunction in small volumes appears clearly.
This behaviour confirms that the node is indeed unphysical and can be
described as a finite volume artifact of our method.

\begin{figure}[tb]
\epsfxsize=65truemm
\centerline{\epsffile[0 130 750 530]{fig4.ps}}
\caption{Scalar glueball wavefunction for 3 values of $\beta$,
using method A. Distances have been rescaled  according to the measured
glueball mass.}
\label{fig:4}
\vspace{-5mm}
\end{figure}

In fact it is clear that the correlation matrix $C$, being real symmetric,
has orthogonal eigenvectors $\psi(r)$.  Since one of them (associated
to the flux state) has no node, all others must have at least one to
ensure orthogonality with the first.

The full wavefunctions of the torelon and glueball, including color
degrees of freedom, are of course orthogonal, but not their radial
projections which we measure \cite{PvB}.  Therefore our initial assumption,
that the eigenvectors of $C(t)$ are pure eigenstates of $H$, is wrong even
though an a posteriori check shows little variation of the eigenvectors
of $C(t)$ with Euclidean time (see Fig.3). This is presumably due to the
closeness in mass of torelon and glueball.

Method {\bf B} recognizes that flux states and glueballs belong to different
sectors.  Within one sector, the radial wavefunctions of various glueball
states
(or flux states) are orthogonal.  But flux state and glueball radial
wavefunctions are not orthogonal to each other.  The correlation matrix (eq.8)
can thus be decomposed into a sum:
\begin{equation}
C(t)  = C_{flux}(t) + C_{glue}(t)
\end{equation}
\vspace{-4mm}
\begin{displaymath}
      = \tilde{R}_{flux} D_{flux}(t) R_{flux}
       + \tilde{R}_{glue} D_{glue}(t) R_{glue}
\end{displaymath}
This decomposition is an ill-posed, unconstrained problem for a single $C(t)$.
We need to require consistency of the eigenvectors $\psi^{(k)}(r)$, and thus
the
rotation matrices $R_{flux}$ and $R_{glue}$, for successive times $t$.
The elements of the torelon wavefunctions $\psi^{(k)}_{flux}(r_i)$ should be
adjusted to minimize the changes in the eigenvectors of $C_{glue}(t)$ with $t$.

This minimization problem is solved iteratively.  At each iteration each
element
of each torelon wavefunction is adjusted. We considered one torelon state, or
two orthogonal ones.  One measure of the mismatch between eigenvectors of
$C_{glue}(0)$ and $C_{glue}(t)$ was
$ \sum_{ij} ( [ C_{glue}(0), C_{glue}(t) ]_{ij} )^2 $, where $[, ]$ is the
commutator.  Another one was
$ \sum_i ( \psi^{(1)}_{glue}(r_i)(t=0) - \psi^{(1)}_{glue}(r_i)(t) )^2 $
where $\psi^{(1)}$ is the eigenvector corresponding to the largest eigenvalue
of $C_{glue}$.  Results depended mildly on the measure chosen.

The effectiveness of this method at removing flux state contributions can be
assessed from Fig.5, where we show the resulting scalar and tensor glueball
wavefunctions at $\beta = 2.5$.  The tensor wavefunction remains positive
and the scalar one barely changes sign.  Including
torelon excited states (only the ground state was considered here) would
further improve the results.  The associated scalar and tensor torelon
wavefunctions, on the same figure, show the expected flatness.

The wavefunctions we obtain are in reasonable agreement with those of Ref.1.
In particular the size of the tensor glueball is clearly much greater than
that of the scalar.  A more systematic analysis along the lines of
\cite{ASK} is under way.

\nopagebreak[3]
\section{Finite temperature}

We are able to extract wavefunctions from very small time separations.
Even the matrix of fluctuations of $W$, corresponding
to a correlation matrix (7) with time separation 0, gives us rather
accurate information (see eg. Fig.3).  Therefore we can study glueball and
flux state propagators in the Euclidean time direction also at finite
temperature. This is in contrast with previous finite temperature studies
of screening masses, which considered propagation along a spatial direction.

To compare with zero-temperature, we studied a deconfined $16^3$ by 4 lattice
at $\beta = 2.5$.  The temperature is around 2 $T_c$.
First results indicate that, surprisingly, no significant changes
occur.  Fig.6 shows the scalar and tensor wavefunctions obtained through the
same analysis as in Fig.5.  Glueball masses appear similarly unaffected.

\begin{figure}[tb]
\epsfxsize=65truemm
\centerline{\epsffile[0 130 750 530]{fig5.ps}}
\caption{Scalar and tensor glueball wavefunctions at $\beta = 2.5$,
using method B. Dotted curves show the scalar and tensor torelon
wavefunctions.}
\label{fig:5}
\end{figure}

\begin{figure}[htb]
\epsfxsize=65truemm
\centerline{\epsffile[0 130 750 530]{fig6.ps}}
\caption{Same as Fig.5, in the deconfined phase.}
\label{fig:6}
\vspace{-5mm}
\end{figure}

This situation is reminiscent of chiral symmetry breaking, where the scalar
$\sigma$ particle remains unaffected by the chiral phase transition, while
the $\pi$ becomes a Goldstone boson.  Here the $Z(N)$ symmetry is broken
in the deconfined phase, and the glueball, a $Z(N)$ scalar, remains unaffected.
It is difficult to go beyond an analogy however, because the broken
$Z(N)$ symmetry is discrete, and an ordinary Goldstone boson cannot exist.

\nopagebreak[3]


\begin{thebibliography}{9}
\bibitem{PvB} This point was strongly, and rightly, emphasized
by P. van Baal.
\nopagebreak[3]
\end{thebibliography}
\end{document}